\documentclass[preprint,showpacs,preprintnumbers,amsmath, 
amssymb,overcite]{revtex4} 
\usepackage{graphicx}
\usepackage{dcolumn}
\usepackage{bm}
\begin{document} 
 
\title{Spectroscopic implications from the combined analysis 
of processes with pseudoscalar mesons\footnote{Supported by the 
Votruba-Blokhintsev Program for Cooperation of the Czech Republic  
with JINR (Dubna), the Grant Agency of the Czech Republic  
(Grant No.202/08/0984), the Slovak Scientific Grant Agency  
(Grant VEGA No.2/0034/09), and the Bogoliubov-Infeld Program for 
Cooperation of Poland with JINR (Dubna).}
\footnote{Talk given at XIII International Conference 
{\it Selected Problems of Modern Theoretical Physics},  
Bogoliubov Laboratory of Theoretical Physics, JINR, Dubna, Russia,  
June 23-27, 2008.}} 
\author{Yu.S.~Surovtsev\footnote{E-mail address: surovcev@thsun1.jinr.ru}} 
\affiliation 
{Bogoliubov Laboratory of Theoretical Physics, JINR, 
Dubna, Russia} 
\author{P.~Byd\v{z}ovsk\'y\footnote{E-mail address: bydz@ujf.cas.cz}} 
\affiliation 
{Nuclear Physics Institute, ASCR, 
\v{R}e\v{z} near Prague, Czech Republic} 
\author{M.~Nagy\footnote{E-mail address: fyzinami@unix.savba.sk}} 
\affiliation 
{Institute of Physics, SAS, Bratislava, 
Slovakia} 
 
\date{19. 11. 2008} 
\begin{abstract} 
In the analysis a status and parameters of the scalar, vector, and  
tensor mesonic resonances are obtained and compared with other  
results. Possible classification of the resonance states in terms  
of the SU(3) multiplets is discussed.  
\end{abstract} 
 
\maketitle 
 
{\bf Outline:} 
\begin{itemize} 
\item Motivation  
\item Method of analysis 
\item Analysis of the isoscalar-scalar sector 
\item Analysis of the isovector $P$-wave of $\pi\pi$ scattering 
\item Analysis of the isoscalar-tensor sector  
\item Spectroscopic implications from the analysis 
\end{itemize} 
 
\section{Motivation} 
The spectroscopy of light mesons plays an important role in 
understanding the strong interactions at low energies.  
Among possibilities to study the spectrum of light mesons,  
analysis of the $\pi\pi$ interaction is particularly useful and,  
therefore, it has always been an object of continuous theoretical  
and experimental investigation \cite{PDG08}. 
Here, we present results of the coupled-channel analysis of data  
on processes $\pi\pi\to\pi\pi,K\overline{K},\eta\eta,\eta\eta^\prime$  
in the channels with $I^GJ^{PC}=0^+0^{++}$ and $0^+2^{++}$ and on the 
$\pi\pi$ scattering in the channel with $1^+1^{--}$. 
 
The scalar sector is problematic up to now especially as to an 
assignment of the discovered mesonic states to quark-model 
configurations in spite of a big amount of work devoted to these 
problems (see, {\it e.g.}, Ref.~\cite{Ani06} and references therein).  
An exceptional interest to this sector is supported by the fact  
that there, possibly indeed, we deal with a glueball $f_0(1500)$  
(see, {\it e.g.}, Ref.~\cite{PDG08,Ams96}). 
 
Investigation of vector mesons is up-to-date subject due to their 
role in forming the electromagnetic structure of particles and 
because our knowledge about these mesons is still too incomplete 
({\it e.g.}, in the Particle Data Group tables \cite{PDG08} (PDG)  
the mass of $\rho(1450)$ is ranging from 1250 to 1582 MeV). 
 
In the tensor sector, among the thirteen discussed resonances,  
the nine states ($f_2(1430)$, $f_2(1565)$, $f_2(1640)$, $f_2(1810)$, 
$f_2(1910)$, $f_2(2000)$, $f_2(2020)$, $f_2(2150)$, $f_2(2220)$) 
must be confirmed in various experiments and analyses. For example,  
in the analysis of $p\overline{p}\to\pi\pi,\eta\eta,\eta\eta^\prime$,  
five resonances -- $f_2(1920)$, $f_2(2000)$, $f_2(2020)$, $f_2(2240)$  
and $f_2(2300)$ -- have been obtained, one of which, $f_2(2000)$, is  
a candidate for the glueball \cite{Ani05}. 
 
In our analysis, we have used both {\it a model-independent  
method} \cite{KMS96}, based on the first principles (analyticity and  
unitarity) directly applied to analysis of experimental data,  
and {\it the multichannel Breit--Wigner forms}.  
The former approach permits us to introduce no theoretical  
prejudice to extracted parameters of resonances, however, it  
is limited with the possibility to use only three coupled  
channels. Therefore, in more general cases, one has to use, {\it e.g.},   
the Breit--Wigner approach. Considering the obtained disposition  
of resonance poles on the Riemann surface, obtained coupling  
constants with channels, and resonance masses we draw particular  
conclusions about nature of the investigated states. 
 
\section{Method of analysis} 
 
In both methods of analysis, we parametrized the $S$-matrix 
elements $S_{\alpha\beta}$ where $\alpha,\beta =1,2,\cdots,n$ 
denote channels, using the Le Couteur-Newton relations \cite{LeCou}. 
This relations express the $S$-matrix elements of all coupled 
processes in terms of the Jost matrix determinant 
$d(k_1,\cdots,k_n)$ that is a real analytic function with the only 
square-root branch-points at the channel momenta $k_{\alpha}=0$. 
 
In the model-independent approach, the $S$-matrix is determined on 
the 4- and 8-sheeted Riemann surfaces for the 2- and 3-channel 
cases, respectively. The matrix elements $S_{\alpha\beta}$ have 
the right-hand cuts along the real axis of the $s$ complex plane 
($s$ is the invariant total energy squared), starting at the 
coupled-channels thresholds $s_i$ ($i=1,2,3$), and the left-hand 
cuts related to the crossed channels.  
The Riemann-surface sheets are numbered according to the signs of 
analytic continuations of the channel momenta  
$k_i=\sqrt{s-s_i}/{2}~~~~(i=1,2,3)$, as shown in Table~\ref{ksigns}. 
%
%
\begin{table}[htb!] 
\caption{Signs of channel momenta on the eight sheets of the Rieman  
surface in the 3-channel case.} 
\label{ksigns} 
\begin{center} 
\begin{ruledtabular} 
{\small  
\begin{tabular}{ccccccccc} 
{sheet:} & I & II & III & IV & V & VI & VII & VIII \\ \hline 
Im $k_1$ & $+$ & $-$ & $-$ & $+$ & $+$ & $-$ & $-$ & $+$ \\ 
Im $k_2$ & $+$ & $+$ & $-$ & $-$ & $-$ & $-$ & $+$ & $+$\\ 
Im $k_3$ & $+$ & $+$ & $+$ & $+$ & $-$ & $-$ & $-$ & $-$ 
\end{tabular}} 
\end{ruledtabular} 
\end{center} 
\end{table} 
 
The model-independent method which essentially utilizes an 
uniformizing variable can be used only for the 2-channel case and 
under some conditions for the 3-channel one. Only in these cases 
we obtain a simple symmetric (easily interpreted) picture of the 
resonance poles and zeros of the $S$-matrix on an uniformization 
plane. The important branch points, corresponding to the 
thresholds of the coupled channels and to the crossing ones, are 
taken into account in the uniformizing variable. 
 
The resonance representations on the Riemann surfaces are obtained 
with the help of formulas from Ref.~\cite{KMS96}, expressing analytic  
continuations of the $S$-matrix elements to unphysical sheets in terms of  
those on sheet I that have only the zeros of resonances (beyond the  
real axis), at least, around the physical region. Then, starting from  
the resonance zeros on sheet I, one can obtain an arrangement of poles  
and zeros of resonance on the whole Riemann surface.  
 
In the 2-channel case, we obtain \underline{three types of resonances} 
described by a pair of conjugate zeros on sheet I: ({\bf a}) in 
$S_{11}$, ({\bf b}) in $S_{22}$, ({\bf c}) in each of $S_{11}$ and 
$S_{22}$. 
 
In the 3-channel case, we obtain \underline{seven types of resonances} 
corresponding to seven possible situations when there are resonance 
zeros on sheet I only in $S_{11}$ -- ({\bf a}); ~~$S_{22}$ -- 
({\bf b}); ~~$S_{33}$ -- ({\bf c}); ~~$S_{11}$ and $S_{22}$ -- 
({\bf d}); ~~$S_{22}$ and $S_{33}$ -- ({\bf e}); ~~$S_{11}$ and 
$S_{33}$ -- ({\bf f}); and ~~$S_{11}$, $S_{22}$, and $S_{33}$ -- ({\bf g}).\\ 
A resonance of every type is represented by a pair of 
complex-conjugate clusters (of poles and zeros on the Riemann 
surface). Note that whereas the cases ({\bf a}), ({\bf b}) and ({\bf 
c}) can be simply related to the representation of resonances by 
the Breit-Wigner forms, the cases ({\bf d}), ({\bf e}), ({\bf f}) and 
({\bf g}) are practically lost at that description. The cluster 
type is related to the nature of state. For example, if we 
consider the $\pi\pi$, $K\overline{K}$, and $\eta\eta$ channels, 
then a resonance which is coupled relatively more strongly to  
the $\pi\pi$ channel than to the $K\overline{K}$ and $\eta\eta$ ones   
is described by the cluster of type ({\bf a}). If the resonance is 
coupled more strongly to the $K\overline{K}$ and $\eta\eta$ 
channels than to the $\pi\pi$ one, then it is represented by  
the cluster of type ({\bf e}) (say, the state with the dominant  
$s{\bar s}$ component).  
The flavour singlet ({\it e.g.}, glueball) must be represented by 
the cluster of type ({\bf g}) (of type ({\bf c}) in the 2-channel 
consideration) as a necessary condition for the ideal case, if 
this state lies above the thresholds of considered channels. 
 
We can distinguish, in a model-independent way, a bound state of 
colourless particles ({\it e.g.}, $K\overline{K}$ molecule) and a 
$q{\bar q}$ bound state. Just as in the 1-channel case, the 
existence of the particle bound-state means the presence of the pole 
on the real axis under the threshold on the physical sheet, so in 
the 2-channel case, the existence of the particle bound-state in 
channel 2 ($K\overline{K}$ molecule) that, however, can decay into 
channel 1 ($\pi\pi$ decay), would imply the presence of a pair of 
complex conjugate poles on sheet II under the second-channel 
threshold without the corresponding shifted pair of poles on sheet III. 
 
In the 3-channel case, the bound-state in channel 3 ($\eta\eta$) 
that, however, can decay into channels 1 ($\pi\pi$ decay) and 2 
($K\overline{K}$ decay), is represented by the pair of complex 
conjugate poles on sheet II and by shifted poles on sheet III 
under the $\eta\eta$ threshold without the corresponding poles on 
sheets VI and VII. This test \cite{KMS96,MPe93} is a multichannel analogue  
of the known Castillejo--Dalitz--Dyson poles in the one-channel case.  
According to this test, earlier in Ref.~\cite{KMS96}, the interpretation  
of the $f_0(980)$ state as the $K\overline{K}$ molecule has been  
rejected because this state is represented by the cluster of type  
({\bf a}) in the 2-channel analysis of processes  
$\pi\pi\to\pi\pi,K\overline{K}$ and, therefore, it does not satisfy  
the necessary condition to be the $K\overline{K}$ molecule. 
 
\section{Analysis of the isoscalar-scalar sector } 
 
Considering the $S$-waves of processes  
$\pi\pi\to\pi\pi,K\overline{K},\eta\eta,\eta\eta^\prime$ in 
{\it the model-independent method}, we performed two variants of the 
3-channel analysis:\\ 
\underline{variant I}: the combined analysis of 
$\pi\pi\to\pi\pi,K\overline{K},\eta\eta\,$;\\ 
\underline{variant II}: analysis of 
$\pi\pi\to\pi\pi,K\overline{K},\eta\eta^\prime$.\\ 
Influence of the $\eta\eta^\prime$-channel in variant I and  
the $\eta\eta$-channel in variant II are taken into account  
via the background. Here, the left-hand cuts are neglected in the 
Riemann-surface structure assuming that contributions on these cuts  
are also included in the background. 
 
Under neglecting the $\pi\pi$-threshold branch 
point (however, unitarity on the $\pi\pi$-cut is taken into 
account), the uniformizing variable is  
\begin{equation} 
w=\frac{k_2+k_3}{\sqrt{m_\eta^2-m_K^2}}~~~~ {\rm for~ variant~ I}, 
\end{equation} 
and 
\begin{equation} 
w^\prime=\frac{k_2^\prime+k_3^\prime} 
{\sqrt{\frac{1}{4}(m_\eta+m_{\eta^\prime})^2-m_K^2}}~~~~ 
{\rm for~ variant~ II}. 
\end{equation} 
The quantities related to variant II are primed.  
 
On the $w$-plane, the Le Couteur-Newton relations are \footnote{ 
Other authors have also used the parameterizations with the Jost 
functions in analyzing the $S$-wave $\pi\pi$ scattering in the 
one-channel approach \cite{Boh80} and in the two-channel one \cite{MPe93}.} 
\begin{eqnarray} 
S_{11}=\frac{d^* (-w^*)}{d(w)},~~~~~~~~  
S_{22}=\frac{d(-w^{-1})}{d(w)},~~~~~~~~ 
S_{33}=\frac{d(w^{-1})}{d(w)},\\ 
S_{11}S_{22}-S_{12}^2=\frac{d^*({w^*}^{-1})}{d(w)},~~~~~~~~ 
S_{11}S_{33}-S_{13}^2=\frac{d^*(-{w^*}^{-1})}{d(w)}\,,  
\end{eqnarray} 
where the $d$-function is assumed in the form 
\begin{equation} 
d=d_B d_{res}, 
\end{equation} 
and the resonance part is 
\begin{equation} 
d_{res}(w)=w^{-\frac{M}{2}}\prod_{r=1}^{M}(w+w_{r}^*) 
\end{equation} 
with $M$ the number of resonance zeros. The background part is taken as  
\begin{equation} 
d_B=\mbox{exp}[-i\sum_{n=1}^{3}\frac{k_n}{m_n}(\alpha_n+i\beta_n)], 
\end{equation} 
where  
\begin{eqnarray} 
\alpha_n=a_{n1}+a_{n\sigma}\frac{s-s_\sigma}{s_\sigma}\theta(s-s_\sigma)+ 
a_{nv}\frac{s-s_v}{s_v}\theta(s-s_v),\\ 
\beta_n=b_{n1}+b_{n\sigma}\frac{s-s_\sigma}{s_\sigma}\theta(s-s_\sigma)+ 
b_{nv}\frac{s-s_v}{s_v}\theta(s-s_v) 
\end{eqnarray}  
with $s_\sigma$ the $\sigma\sigma$ threshold and $s_v$ a combined threshold  
of many opened channels in the vicinity of 1.5 GeV 
({\it e.g.}, $\eta\eta^{\prime},~\rho\rho,~\omega\omega$). 
 
In variant II, the terms 
\begin{equation} 
a_{n\eta}^\prime\frac{s-4m_\eta^2}{4m_\eta^2}\theta(s-4m_\eta^2)~~~~~ 
{\rm and}~~~~~b_{n\eta}^\prime\frac{s-4m_\eta^2}{4m_\eta^2}\theta(s-4m_\eta^2) 
\end{equation} 
should be added to $\alpha^\prime_n$ and $\beta^\prime_n$ to 
account for an influence of the $\eta\eta$-channel. 
 
As the data, we use the results 
of phase analyses given for phase shifts of the amplitudes 
$\delta_{ab}$ and for moduli of the $S$-matrix elements $\eta_{ab}=|S_{ab}|$ 
($a,b=$1-$\pi\pi$,~2-$K\overline{K}$,~3-$\eta\eta$ or $\eta\eta^{\prime}$): 
\begin{equation} 
S_{aa}=\eta_{aa}e^{2i\delta_{aa}},~~~~~S_{ab}=\eta_{ab}e^{i\phi_{ab}}. 
\end{equation} 
If below the $\eta\eta$-threshold there is the 2-channel unitarity, then the 
relations 
\begin{equation} 
\eta_{11}=\eta_{22}, ~~ \eta_{12}=(1-{\eta_{11}}^2)^{1/2},~~ 
\phi_{12}=\delta_{11}+\delta_{22} 
\end{equation} 
are fulfilled in this energy region. 

The $\pi\pi$ scattering data, which range from the threshold up to 1.89 GeV,   
are taken from Ref.~\cite{Hya73,expd1}\footnote{Note that there are  
alternative data, e.g., one of the solutions of the phase analysis 
in Ref. \cite{Grayer} and the recent phase analysis in Ref. \cite{Kaminski_data} 
which are in accordance with each other, but which differ from those 
used here, especially in the $f_0(980)$ region of energy. Analysis with these 
data should be performed separately. This work is in progress.}. 
For $\pi\pi\to K\overline{K}$, 
practically all the accessible data are used \cite{expd2}. For 
$\pi\pi\to\eta\eta$, we used data for $|S_{13}|^2$ from the 
threshold to 1.72 GeV \cite{expd3}. For $\pi\pi\to\eta\eta^\prime$, 
the data for $|S_{13}|^2$ from the threshold to 1.813 GeV are 
taken from Ref.~\cite{expd4}. We included all the five resonances  
discussed below 1.9 GeV. 
 
In \underline{variant I}, 
we got satisfactory description: for the $\pi\pi$ scattering, 
$\chi^2/\mbox{NDF}\approx1.35$; for $\pi\pi\to K\overline{K}$, 
$\chi^2/\mbox{NDF}\approx 1.77$; for $\pi\pi\to\eta\eta$, 
$\chi^2/\mbox{N.exp.points}\approx 0.86$. 
The total $\chi^2/\mbox{NDF}$ is $345.603/(301-40)\approx1.32$.  
From possible resonance representations by pole-clusters, the 
analysis selects the following one: the $f_0 (600)$ is described 
by the cluster of type ({\bf a}); $f_0 (1370)$, type ({\bf c}); 
$f_0 (1500)$, type ({\bf g}); $f_0 (1710)$, type ({\bf b}); and the 
$f_0 (980)$ is represented only by the pole on sheet II and 
shifted pole on sheet III in both variants. 
The background parameters are: $a_{11}=0.2006$, 
$a_{1\sigma}=0.0146$, $a_{1v}=0$, $b_{11}=0$, 
$b_{1\sigma}=-0.01025$, $b_{1v}=0.0542$, $a_{21}=-0.6986$, 
$a_{2\sigma}=-1.4207$, $a_{2v}=-5.958$, $b_{21}=0.047$, 
$b_{2\sigma}=0$, $b_{2v}=6.888$, $b_{31}=0.6511$, 
$b_{3\sigma}=0.3404$, $b_{3v}=0$; $s_\sigma=1.638~{\rm GeV}^2$, 
$s_v=2.084~{\rm GeV}^2$. 
 
In \underline{variant II}, we got the following description: for 
the $\pi\pi$ scattering $\chi^2/\mbox{NDF}\approx 1.0$! 
for~$\pi\pi\to K\overline{K}$  $\chi^2/\mbox{NDF}\approx 1.62$; 
for~$\pi\pi\to\eta\eta^\prime$  $\chi^2/\mbox{N.exp.points}\approx 
0.36$. The total $\chi^2/\mbox{NDF}$ is $282.682/(293-38)\approx 
1.11$! In this case, the $f_0 (600)$ is described by the cluster 
of type ({\bf a}$^\prime$); $f_0 (1370)$, type ({\bf b}$^\prime$); 
$f_0 (1500)$, type ({\bf d}$^\prime$); and $f_0(1710)$, type ({\bf 
c}$^\prime$). The background parameters are: 
$a_{11}^\prime=0.0111$, $a_{1\eta}^\prime=-0.058$, 
$a_{1\sigma}^\prime=0$, $a_{1v}^\prime=0.0954$, 
$b_{11}^\prime=b_{1\eta}^\prime=b_{1\sigma}^\prime=0$, 
$b_{1v}^\prime=0.047$, $a_{21}^\prime=-3.439$, 
$a_{2\eta}^\prime=-0.4851$, $a_{2\sigma}^\prime=1.7622$, 
$a_{2v}^\prime=-5.158$, $b_{21}^\prime=0$, 
$b_{2\eta}^\prime=-0.7524$, $b_{2\sigma}^\prime=2.6658$, 
$b_{2v}^\prime=1.836$, $b_{31}^\prime=0.5545$, 
$s_\sigma=1.638~{\rm GeV}^2$, $s_v=2.126~{\rm GeV}^2$. 
 
\begin{figure}[htb] 
\begin{center} 
\includegraphics[width=0.85\textwidth]{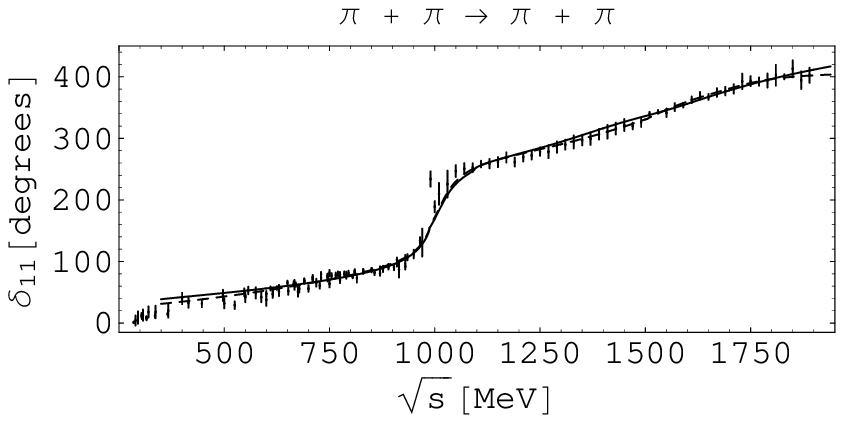} 
\includegraphics[width=0.85\textwidth]{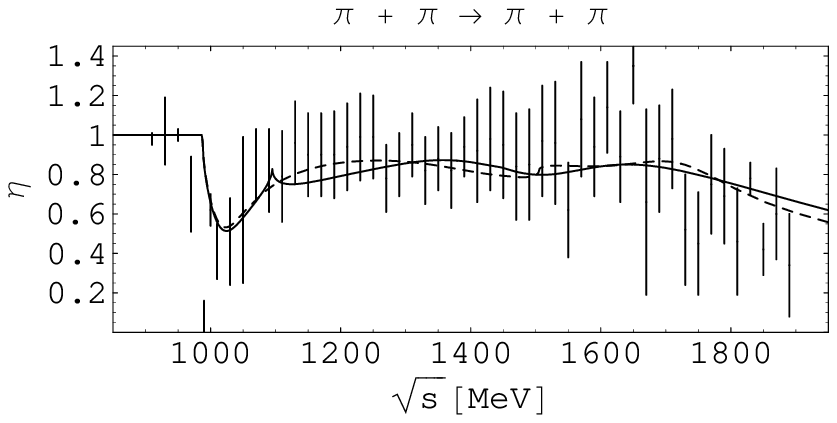} 
\caption{The phase shift and module of the $S$-matrix element  
in the $S$-wave $\pi\pi$-scattering. The solid curve corresponds to  
variant I and the dashed curve to variant II.} 
\label{fig:s-wave1} 
\end{center} 
\end{figure} 
\begin{figure}[htb] 
\begin{center} 
\includegraphics[width=0.85\textwidth]{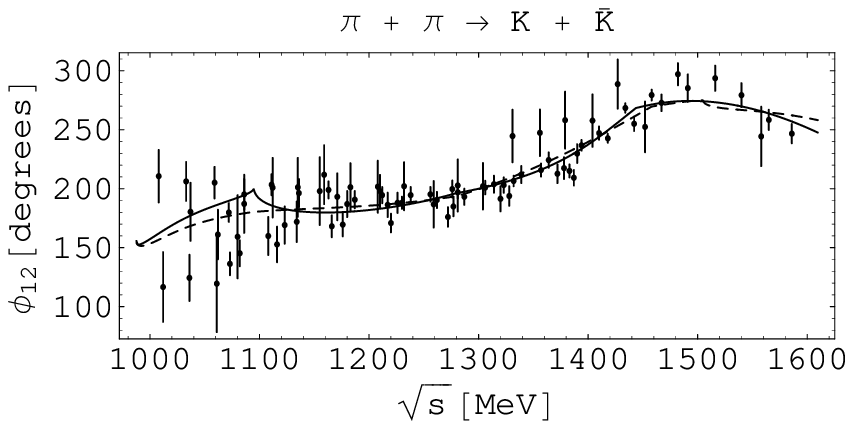} 
\includegraphics[width=0.85\textwidth]{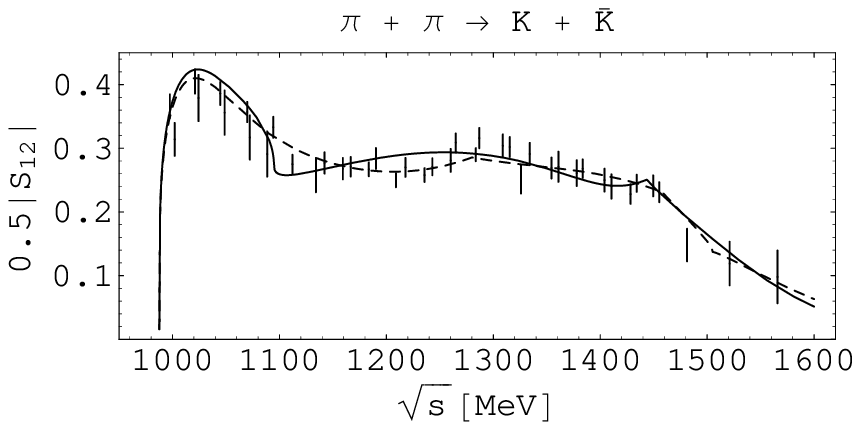} 
\caption{The phase shift and module of the $S$-matrix element  
in $S$-wave of $\pi\pi\to K\overline{K}$. The solid curve corresponds to  
variant I and the dashed curve to variant II.} 
\label{fig:s-wave2} 
\end{center} 
\end{figure} 
\begin{figure}[htb] 
\begin{center} 
\includegraphics[width=0.85\textwidth]{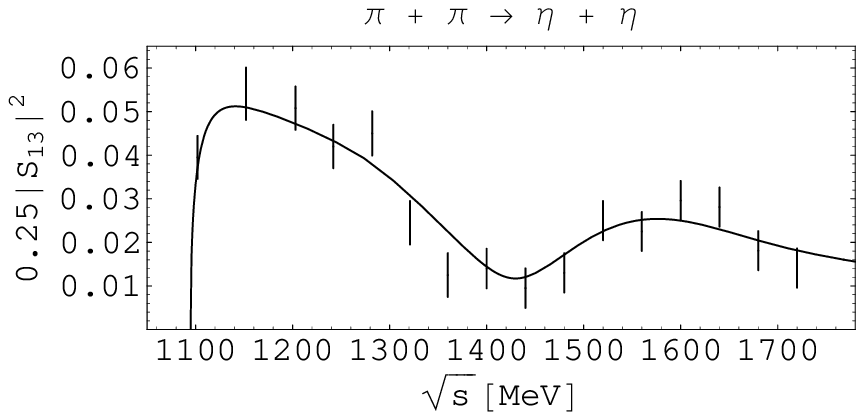} 
\includegraphics[width=0.85\textwidth]{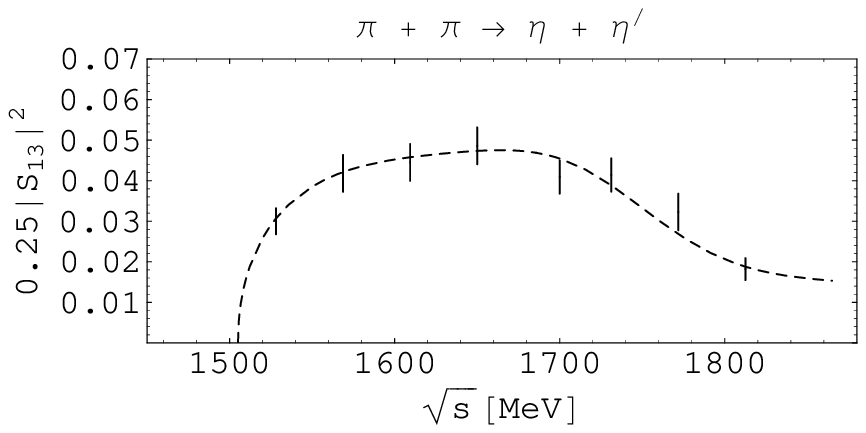} 
\caption{The squared modules of the $\pi\pi\to\eta\eta$ (upper figure) and 
$\pi\pi\to\eta\eta^\prime$ (lower figure) $S$-wave matrix elements.} 
\label{fig:s-wave3} 
\end{center} 
\end{figure} 
In Figures~\ref{fig:s-wave1}-\ref{fig:s-wave3}, we show results  
of fitting to the experimental data and in Table~\ref{tab:clusters}  
we indicate the obtained pole clusters for resonances on the eight  
sheets of the complex energy plane $\sqrt{s}$, on which the 3-channel  
$S$-matrix is determined ($\!\sqrt{s_r}={\rm E}_r-i\Gamma_r\!$). 
%
%
\begin{table}[htb] 
\caption{Pole clusters for the $f_0$-resonances in variants I and II.} 
\label{tab:clusters} 
\begin{center} 
\begin{ruledtabular} 
{\small  
\begin{tabular}{ccccccccc} 
\multicolumn{2}{c}{Sheet} & II & III & IV & V & VI & VII & VIII \\ \hline  
\multicolumn{9}{c}{variant I}\\ \hline 
{$\!\!f_0 (600)\!\!$} & {$\!\!{\rm E}_r\!\!$} & 598.2$\pm$13 & 
585.8$\!\pm\!$14 & {} & {} & 505.8$\!\pm\!$16 & 518.2$\!\pm\!$15 & {} \\ {} & 
{$\!\Gamma_r\!$} & 583$\!\pm\!$18 & 583$\!\pm\!$18 & {} &{} & 583$\!\pm\!$18 & 
583$\!\pm\!$18 & {}\\ \hline {$\!f_0(980)\!$} & {$\!{\rm E}_r\!$} & 
1013.1$\!\pm\!$4 & 983.6$\!\pm\!$9 & {} & {} & {} & {} & {}\\ {} & 
{$\!\Gamma_r\!$} & 34.1$\!\pm\!$6 & 57.4$\!\pm\!$10 & {} & {} & {} & {} & {} 
\\ \hline {$\!f_0 (1370)\!$} & {$\!{\rm E}_r\!$} 
& {} & {} & {} & 1398.2$\!\pm\!$16 & 1398.2$\!\pm\!$18 & 1398.2$\!\pm\!$18 
& 1398.2$\!\pm\!$13 \\ 
{} & {$\!\Gamma_r\!$} 
& {} & {} & {} & 287.4$\!\pm\!$17 & 270.6$\!\pm\!$15 & {155$\!\pm\!$9} 
& {171.8$\!\pm\!$7} \\ 
\hline {$\!f_0 (1500)\!$} & {$\!{\rm E}_r\!$} & 1502.6$\!\pm\!$11 & 
1479.5$\!\pm\!$13 & 1502.6$\!\pm\!$12 & 1496.7$\!\pm\!$12 & 1498$\!\pm\!$16 & 
1496.8$\!\pm\!$12 & 1502.6$\!\pm\!$10 \\ {} & {$\!\Gamma_r\!$} 
& 357.1$\!\pm\!$15 & 
139.4$\!\pm\!$12 & 238.7$\!\pm\!$13 & 139.9$\!\pm\!$14 & 191.2$\!\pm\!$17 & 
87.36$\!\pm\!$11 & 356.5$\!\pm\!$14 \\ \hline {$\!f_0 (1710)\!$} 
& {$\!{\rm E}_r\!$} 
& {} & 1708.2$\!\pm\!$12 & 1708.2$\!\pm\!$10 & 1708.2$\!\pm\!$13 
& 1708.2$\!\pm\!$15 & {} & {}\\ 
{} & {$\!\Gamma_r\!$} & {} & 142.3$\!\pm\!$9 & 160.3$\!\pm\!$8 
& 323.3$\!\pm\!$14 & 305.3$\!\pm\!$13 & {} & {}\\ \hline 
\multicolumn{9}{c}{variant II}\\ \hline 
{$\!f_0 (600)\!$} & {$\!{\rm E}_r\!$} & 616.5$\!\pm\!$8 & 621.8$\!\pm\!$10 &  
{} & {} & 598.3$\!\pm\!$11 & 593$\!\pm\!$12 & {} \\  
{} & {$\!\Gamma_r\!$} & 563$\!\pm\!$11 & 563$\!\pm\!$12 & {} &{} &  
563$\!\pm\!$14 & 563$\!\pm\!$13 & {}\\ \hline  
{$\!f_0(980)\!$} & {$\!{\rm E}_r\!$} & 1009.3$\!\pm\!$3 & 986$\!\pm\!$6 & 
{} & {} & {} & {} & {}\\ {} & {$\!\Gamma_r\!$} & 32$\!\pm\!$4 & 58$\!\pm\!$5.5 
& {} & {} & {} & {} & {} \\ \hline {$\!f_0 (1370)\!$} & {$\!{\rm E}_r\!$} 
& {} & 1394.3$\!\pm\!$9 & 1394.3$\!\pm\!$11 & 1412.7$\!\pm\!$13 &  
1412.7$\!\pm\!$14 & {} & {} \\ 
{} & {$\!\Gamma_r\!$} & {} & 236.3$\!\pm\!$10 & 255.7$\!\pm\!$12 &  
255.7$\!\pm\!$12 & 236.3$\!\pm\!$19 & {} & {} \\ \hline  
{$\!f_0 (1500)\!$} & {$\!{\rm E}_r\!$} & 1498.3$\!\pm\!$11 & 1502.4$\!\pm\!$9 
& 1498.3$\!\pm\!$12 & 1498.3$\!\pm\!$13 & 1494.6$\!\pm\!$11 & 1498.3$\!\pm\!$14 
& {} 
\\ {} & {$\!\Gamma_r\!$} & 198.8$\!\pm\!$14 & 236.8$\!\pm\!$11 
& 193$\!\pm\!$9 & 198.8$\!\pm\!$11 & 194$\!\pm\!$8 & 193$\!\pm\!$10 & {} \\ 
\hline {$\!f_0 (1710)\!$} & {$\!{\rm E}_r\!$} & {} & {} & {} & 
1726.1$\!\pm\!$12 & 1726.1$\!\pm\!$13 & 1726.1$\!\pm\!$12 & 
1726.1$\!\pm\!$10 \\ 
 & {$\!\Gamma_r\!$} & {} & {} & {} & 140.2$\!\pm\!$9 & 111.6$\!\pm\!$8 &  
84.2$\!\pm\!$8 & 112.8$\!\pm\!$7 
\end{tabular} 
} 
\end{ruledtabular} 
\end{center} 
\end{table} 

The $f_0(1370)$ and $f_0(1710)$ are represented by the pole 
clusters corresponding to states with the dominant $s\bar{s}$ 
component; $f_0(1500)$, with the dominant glueball component. 
 
Note a surprising result obtained for the $f_0(980)$. This state 
lies slightly above the $K\overline{K}$ threshold and is described 
by the pole on sheet II and by the shifted pole on sheet III under 
the $\eta\eta$ threshold without the corresponding poles on sheets 
VI and VII, as it was expected for standard clusters. This 
corresponds to the description of the $\eta\eta$ bound state. 
 
Masses and total widths of states should be calculated from the pole 
positions. If, when calculating these quantities, the resonance part 
of amplitude is taken in the form 
\begin{equation} 
T^{res}=\frac{\sqrt{s_r}\Gamma_{el}}{m_{res}^2-s_r-i\sqrt{s_r}\Gamma_{tot}},  
\end{equation} 
we obtain values of masses and total widths of the 
$f_0$-resonances, presented in Table \ref{tab:mass}. 
%
%
\begin{table}[ht] 
\caption{Masses and total widths of the $f_0$-resonances (all in MeV).} 
\label{tab:mass} 
\begin{center} 
\begin{ruledtabular} 
\begin{tabular}{ccccc} 
{} & \multicolumn{2}{c}{Variant I}& 
\multicolumn{2}{c}{Variant II}\\\hline {State} & $m_{res}$ & 
$\Gamma_{tot}$ & $m_{res}$ & $\Gamma_{tot}$ \\ \hline $f_0 (600)$ 
& 835.3 & 1166 & 834.9 & 1126 \\ \hline $f_0 (980)$ & 1013.7 & 
68.2 & 1009.8 & 64 \\ \hline $f_0 (1370)$ & 1408.7 & 343.6 & 
1417.5 & 511 \\ \hline $f_0 (1500)$ & 1544 & 714 & 1511.4 & 398 \\ 
\hline $f_0 (1710)$ & 1715.7 & 321 & 1729.8 & 225.6 
\end{tabular} 
\end{ruledtabular} 
\end{center} 
\end{table} 
 
\section{Analysis of the isovector $P$-wave of $\pi\pi$ scattering } 
 
In this sector we applied both {\it the model-independent method}  
and {\it multichannel Breit--Wigner forms}.  
We analyzed data in Ref.~\cite{expd5,Hya73}, for the inelasticity parameter 
($\eta$) and phase shift of the $\pi\pi$-scattering amplitude 
($\delta$) ($S(\pi\pi\to\pi\pi)=\eta\exp(2i\delta)$), introducing 
three ($\rho(770)$, $\rho(1250)$ and $\rho(1550-1780)$), four (the 
indicated ones plus $\rho(1860-1910)$) and five (the indicated 
four plus $\rho(1450)$) resonances \cite{SB_NPA08}. 
 
\subsection{The Model-Independent Analysis} 
 
Since in the data for the $P$-wave $\pi\pi$ scattering a deviation 
from elasticity is observed in the near-threshold region of the 
$\omega\pi$ channel, we considered explicitly the thresholds of the 
$\pi\pi$ and $\omega\pi$ channels and the left-hand one at $s=0$ 
in the uniformizing variable: 
\begin{equation} 
v=\frac{(m_\omega+m_{\pi^0})/2~\sqrt{s-4m_{\pi^+}^2} + 
m_{\pi^+}~\sqrt{s-(m_\omega+m_{\pi^0})^2}}{\sqrt{s\left[\left((m_\omega+ 
m_{\pi^0})/2\right)^2-m_{\pi^+}^2\right]}}. 
\end{equation}  
Influence of other channels which couple to the $\pi\pi$ one is  
supposed to be taken into account via the background. 
 
On the $v$-plane, the resonance part of the 2-channel $S$-matrix 
element of $\pi\pi$-scattering $S_{res}$ has no cuts and has the form  
\begin{equation} 
S_{res}=\frac{d(-v^{-1})}{d(v)},  
\end{equation} 
where $d(v)$ represents the contribution of resonances \cite{SB_NPA08}.   
 
The background part is 
\begin{eqnarray} \label{S_bg} 
S_{bg}=\exp\left[2i\left(\sqrt{\frac{s-4m_{\pi^+}^2}{s}}\right)^{3}\left(\alpha_0 
+ \alpha_1~\frac{s-s_1}{s}~\theta(s-s_1) +\right.\right.\nonumber\\ 
\left.\left.\alpha_2~\frac{s-s_2}{s}~\theta(s-s_2)\right)\right]\,, 
\end{eqnarray} 
where $\alpha_i=a_i+ib_i$, $s_1$ is the threshold of 4$\pi$ 
channel noticeable in the $\rho$-like meson decays and $s_2$ is 
the threshold of $\rho2\pi$ channel. Due to allowing for the 
left-hand branch-point at $s=0$ in the $v$-variable, $a_0=b_0=0$. 
Furthermore, $b_1=0$ which is related to the experimental fact 
that the $P$-wave $\pi\pi$ scattering is elastic also above the 
4$\pi$-channel threshold up to about the $\omega\pi^0$ threshold. 
 
\begin{figure}[htb] 
\begin{center} 
\includegraphics[width=0.45\textwidth,angle=-90]{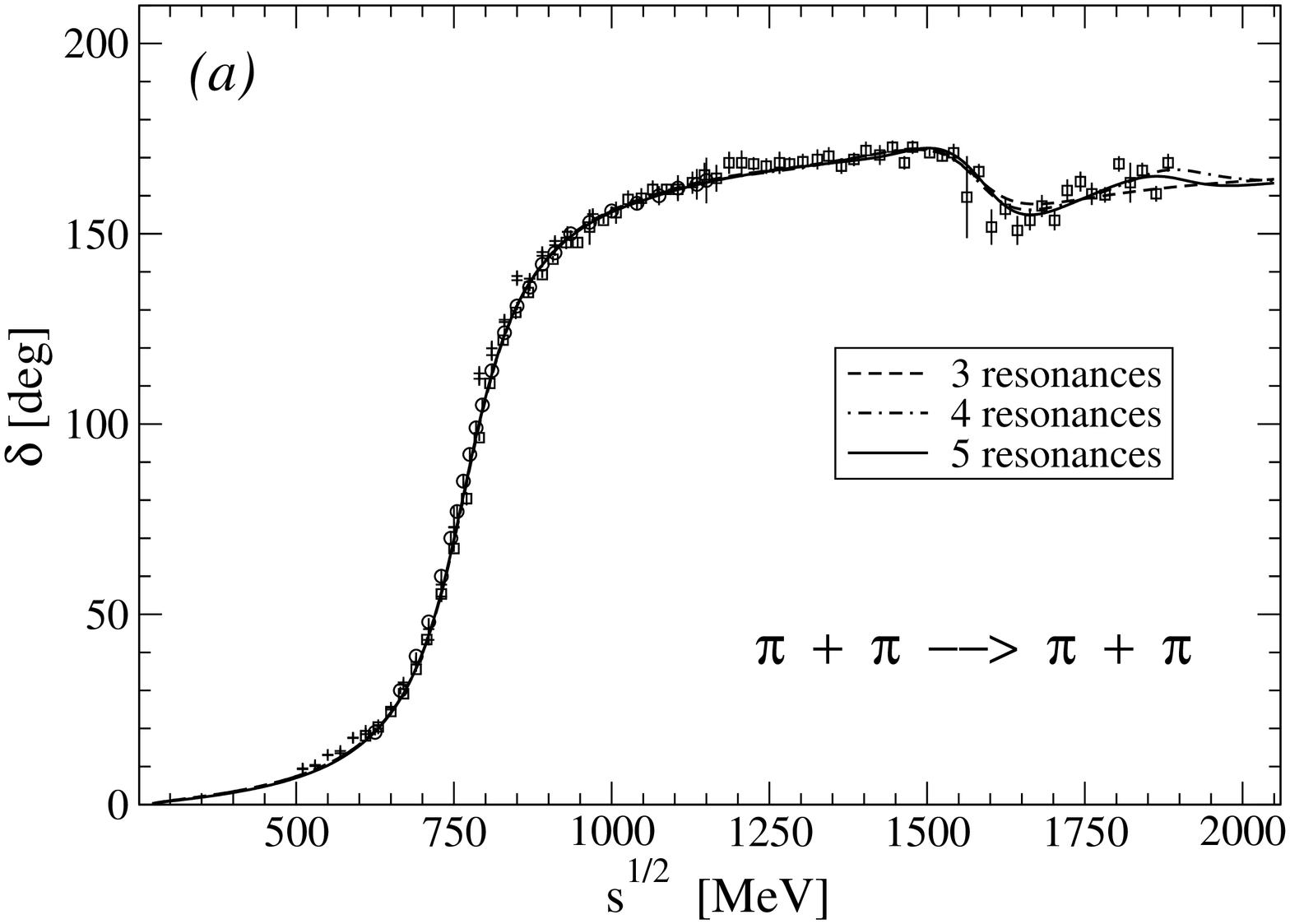} 
\includegraphics[width=0.45\textwidth,angle=-90]{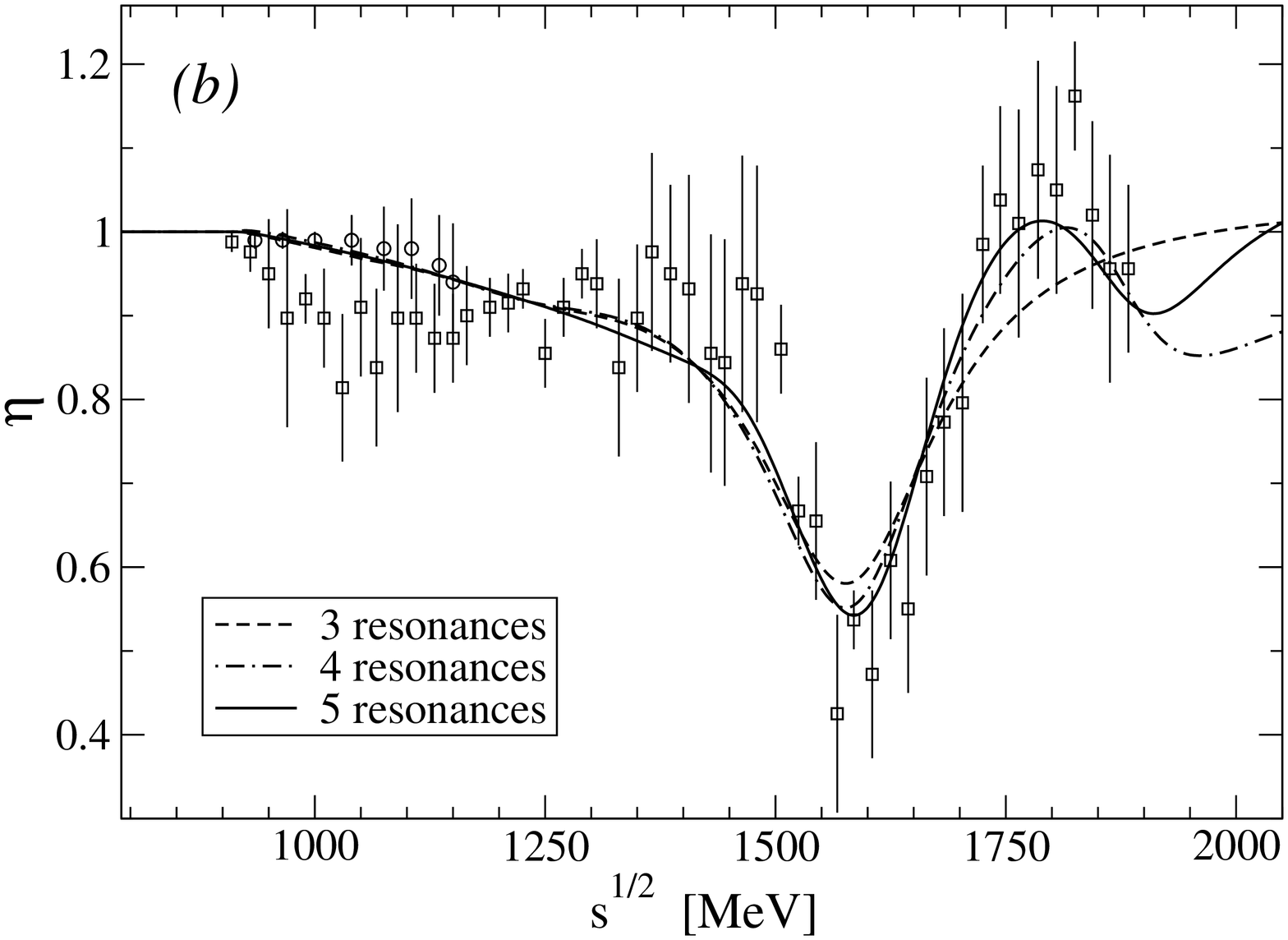} 
\end{center} 
\caption{The phase shift of amplitude and module of the  
$S$-matrix element for the $P$-wave $\pi\pi$-scattering in  
the model-independent approach.} \label{vector} 
\end{figure}

In Figure \ref{vector} we present results of fitting to the data with 
three, four and five resonances. 
We obtained satisfactory description with the total 
$\chi^2/\mbox{NDF}$ equal to $~291.76/(183-15)=1.74$, 
$~278.50/(183-19)=1.70$, and $~266.14/(183-23)=1.66$ for  
the case of three, four and five resonances, respectively.  
 
The $\rho(770)$ is described by the cluster of type ({\bf a})  
and the others by type ({\bf b}). 
The background parameters are: $a_1=0.0093\pm0.0199$, 
$a_2=0.0618\pm0.0305$, and $b_2=-0.0135\pm0.0371$ for the 
three-resonance, $a_1=0.0017\pm0.2118$, $a_2=0.0433\pm0.3552$, and 
$b_2=-0.0044\pm0.4782$ for the four-resonance, and 
$a_1=0.0256\pm0.0186$, $a_2=0.0922\pm0.0335$, and 
$b_2=0.0011\pm0.0478$ for the five-resonance descriptions. The 
positive sign of $b_2$ in the last case is more natural from the 
physical point of view. 
 
Though the description can be considered, 
practically, as the same in all three cases, careful comparison of 
the obtained parameters and energy dependence of the fitted 
quantities suggests that the resonance $\rho(1900)$ is desired and 
that the $\rho(1450)$ might be also included improving slightly 
the description (at all events, its existence does not contradict 
to the data). 
 
In Table \ref{tab:5resMI}, we show the pole clusters of the $\rho$-like 
states on the lower $\sqrt{s}$-half-plane (in MeV) (the conjugate 
poles on the upper half-plane are not shown). 
%
%
\begin{table}[ht] 
\caption{Pole clusters distributed on the sheets II, III, and IV  
for the case with five $\rho$-like resonances. $\sqrt{s_r}$ in MeV  
is given.} 
\label{tab:5resMI} 
\begin{center} 
\begin{ruledtabular} 
\begin{tabular}{cccc} 
&II & III & IV \\ 
\hline 
$\rho(770)$  & $\!765.8\pm 0.6-i(73.3\pm0.4)\!$ & $\!778.2\pm9.1-i(68.9\pm3.9)\!$ 
& {} \\ 
$\rho(1250)$ &  & $\!1251.4\pm11.3-i(130.9\pm9.1)\!$ & 
$\!1251\pm11.1-i(130.5\pm9.2)\!$ \\ 
$\rho(1470)$ &  & $\!1469.4\pm10.6-i(91\pm12.9)\!$ & 
$\!1465.4\pm12.1-i(99.8\pm15.6)\!$ \\ 
$\rho(1600)$ &  & $\!1634\pm20.1-i(144.7\pm23.8)\!$ 
& $\!1592.9\pm7.9-i(73.7\pm11.7)\!$ \\ 
$\rho(1900)$ &  & $\!1882.8\pm24.8-i(112.4\pm25.2)\!$ & 
$\!1893\pm21.9-i(93.4\pm19.9)\!$ 
\end{tabular} 
\end{ruledtabular} 
\end{center} 
\end{table} 
 
Masses and total widths of the obtained $\rho$-states can be 
calculated from the pole positions on sheets II and IV for 
resonances of type ({\bf a}) and ({\bf b}), respectively. The obtained 
values are shown in Table \ref{masses}. 
%
%
\begin{table}[ht] 
\caption{Calculated masses and total widths of the $\rho$-states (all in MeV).} 
\label{masses} 
\begin{center} 
\begin{ruledtabular} 
\begin{tabular}{lrr} 
& $m_{res}$ & $\Gamma_{tot}$\\ 
\hline 
$\rho(770)$  & ~~769.3$\pm$0.6  & ~~146.6$\pm$0.9\\ 
$\rho(1250)$ & ~~1257.8$\pm$11.1 & ~~261$\pm$18.3\\ 
$\rho(1470)$ & ~~1468.8$\pm12.1$ & ~~199.6$\pm$31.2\\ 
$\rho(1600)$ & ~~1594.6$\pm8$ & ~~147.4$\pm$23.4\\ 
$\rho(1900)$ & ~~1895.3$\pm$21.9 & ~~186.8$\pm$39.8 
\end{tabular} 
\end{ruledtabular} 
\end{center} 
\end{table} 
 
\subsection{The Breit--Wigner Analysis} 
 
We used the 5-channel 
Breit--Wigner forms in constructing the Jost matrix determinant 
$d(k_1,\cdots,k_5)$. The resonance poles and zeros in the 
$S$-matrix are generated utilizing the Le~Couteur--Newton relation 
\begin{equation} 
S_{11}=\frac{d(-k_1,\cdots,k_5)}{d(k_1,\cdots,k_5)}\; , 
\end{equation} 
where $k_1$, $k_2$, $k_3$, $k_4$, and $k_5$ are the momenta of 
$\pi\pi$, $\pi^+\pi^-2\pi^0$, $2\pi^+2\pi^-$, $\eta2\pi$, and 
$\omega\pi^0$ channels, respectively. The Jost function is taken as  
\begin{equation} 
d=d_{res}d_{bg}\,, 
\end{equation} 
where the resonance part is 
\begin{equation} 
d_{res}(s)=\prod_{r} 
\left[M_r^2-s-i\sum_{j=1}^5\rho_{rj}^3~R_{rj}~f_{rj}^2\right] 
\end{equation} 
with $\rho_{rj}=k_j(s)/k_j(M_r^2)$ and $f_{rj}^2/M_r$ the partial 
width of a resonance of mass $M_r$. $R_{rj}$ is a 
Blatt--Weisskopf barrier factor: 
\begin{equation} 
R_{rj}=\frac{1+\frac{1}{4}(\sqrt{M_r^2-4m_j^2}~r_{rj})^2} 
{1+\frac{1}{4}(\sqrt{s-4m_j^2}~r_{rj})^2} 
\end{equation} 
with radius $r_{rj}=0.7035$ fm for all resonances in all 
channels as a result of our analysis. Furthermore, we have assumed 
that the widths of resonance decays to $\pi^+\pi^-2\pi^0$ and 
$2(\pi^+\pi^-)$ channels are related each other by relation: 
$f_{r2}=f_{r3}/\sqrt{2}$. This relation is well justified with  
a 5-10\% accuracy, for example, by calculations of the $\rho^0$-meson  
decays in some variant of the chiral model \cite{Ach05}. 
 
The background part of the Jost function is 
\begin{equation} 
d_{bg}=\exp\left[-i\left(\sqrt{\frac{s-4m_{\pi^+}^2}{s}}\right)^3\left(\alpha_0+ 
\alpha_1~\frac{s-s_1}{s}~\theta(s-s_1)\right) \right]\; , 
\end{equation} 
where $\alpha_i=a_i+ib_i$ and $s_1$ is the threshold of the $\rho2\pi$ channel. 
 
\begin{figure}[htb] 
\begin{center} 
\includegraphics[width=0.45\textwidth,angle=-90]{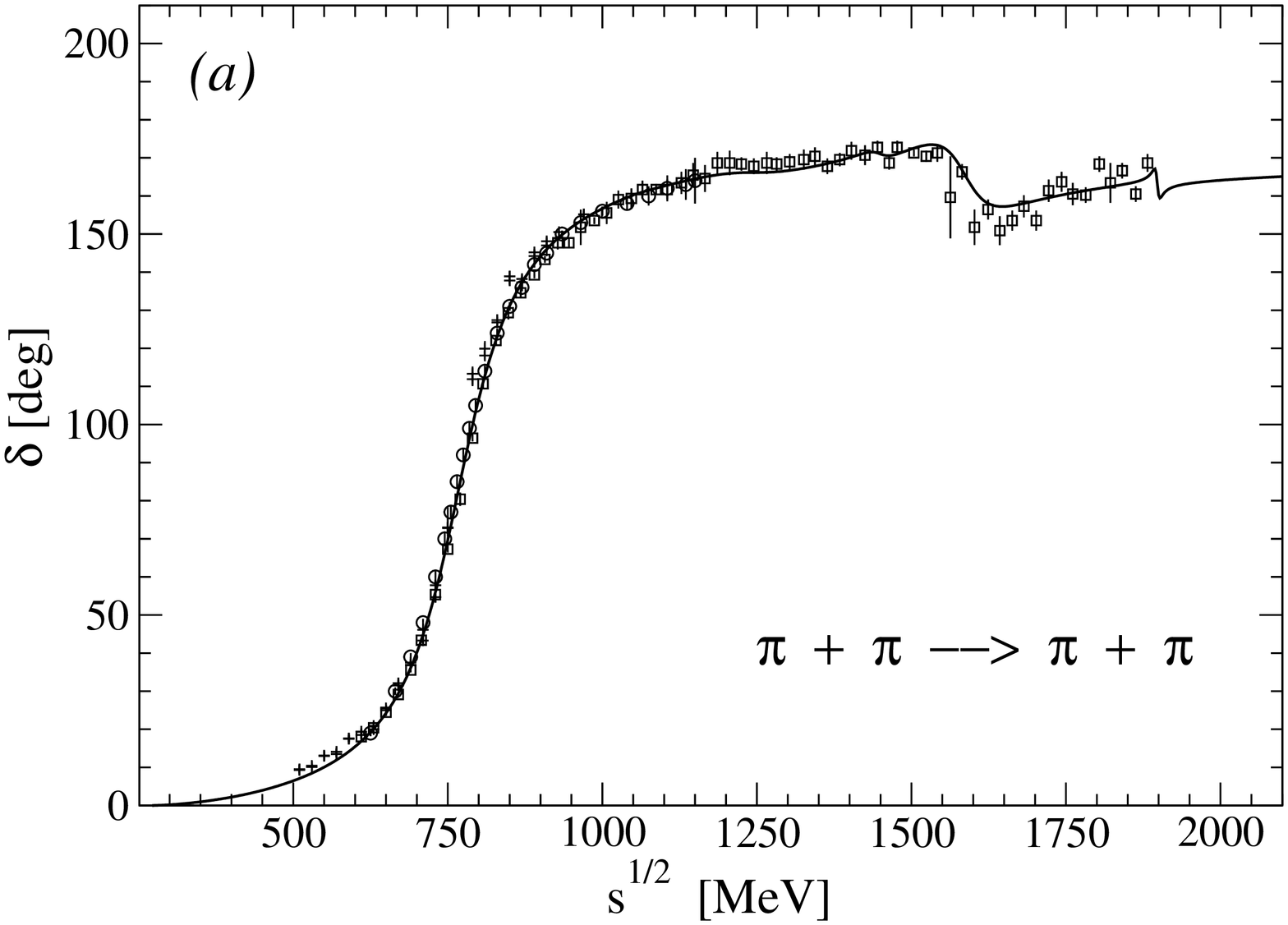} 
\includegraphics[width=0.45\textwidth,angle=-90]{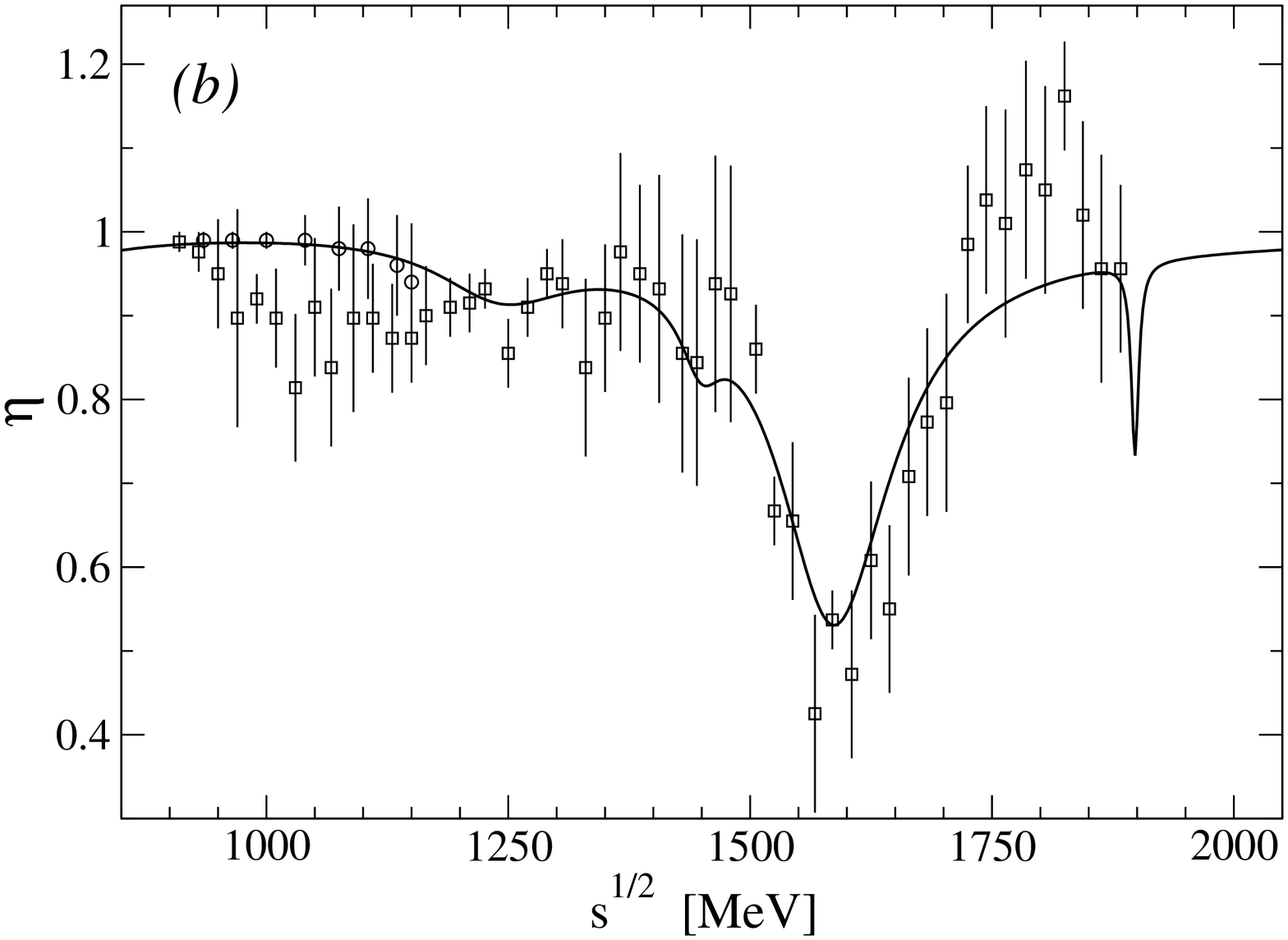} 
\end{center} 
\caption{The phase shift of amplitude and module of the $S$-matrix element  
for the $P$-wave $\pi\pi$-scattering for the case of five resonances  
in the Breit-Wigner approach.} \label{vect-BW} 
\end{figure}

In Figure~\ref{vect-BW}, results of fitting to the data are shown and  
in Table \ref{vparam}, the $\rho$-like resonance parameters are presented.  
We obtained equally reasonable description in all three 
cases: the total $\chi^2/\mbox{NDF}=316.21/(183-17)=1.87$, 
$314.69/(183-22)=1.92$, and $303.10/(183-27)=1.91$ for the case of 
three, four, and five resonances, respectively.  
%
%
\begin{table}[ht] 
\caption{The $\rho$-like resonance parameters in the Breit-Wigner analysis  
(all in MeV).} 
\label{vparam} 
\begin{center} 
\begin{ruledtabular} 
\begin{tabular}{cccccc} 
{State} &{$\rho(770)$}&{$\rho(1250)$}&{$\rho(1450)$}&{$\rho(1600)$}&
{$\rho(1900)$}\\ 
\hline 
~$M$~&777.69$\!\pm\!$0.32&1249.8$\!\pm\!$15.6&1449.9$\!\pm\!$12.2 
&1587.3$\!\pm\!$4.5&1897.8$\!\pm\!$38\\ 
\hline 
$f_{r1}$&343.8$\!\pm\!$0.73&87.7$\!\pm\!$7.4&56.9$\!\pm\!$5.4&248.2$\!\pm\!$5.2 
&47.3$\!\pm\!$12\\ 
\hline 
$f_{r2}$&24.6$\!\pm\!$5.8&186.3$\!\pm\!$39.9&100.1$\!\pm\!$18.7&240.2$\!\pm\!$8.6 
&73.7\\ 
\hline 
$f_{r3}$&34.8$\!\pm\!$8.2&263.5$\!\pm\!$56.5&141.6$\!\pm\!$26.5&339.7$\!\pm\!$12.5 
&104.3\\ 
\hline 
$f_{r4}$&{}&231.8$\!\pm\!$111&141.2$\!\pm\!$98&141.8$\!\pm\!$33& 9 \\ 
\hline 
$f_{r5}$&{}&231$\!\pm\!$115&150$\!\pm\!$95&108.6$\!\pm\!$40.4& 10 \\ 
\hline 
$\Gamma_{tot}$&$\approx$154.3&$>$175&$>$52&$>$168&$>$10 
\end{tabular} 
\end{ruledtabular} 
\end{center} 
\end{table} 
The background parameters for the five-resonance description are:  
$a_0=-0.00121\pm0.0018$, $a_1=-0.1005\pm0.011$, and $b_1=0.0012\pm0.006$.  
The background parameters for the other two cases can be found in  
Ref.~\cite{SB_NPA08}. 
 
In order to look at consistency of the description, we checked if 
the obtained formula for the $\pi\pi$-scattering amplitude gives a 
value of the scattering length consistent with the results of 
other approaches (Table \ref{scatlength}). It seems that the 
satisfactory agreement we obtained is not accidental, because in 
the energy region from the $\pi\pi$ threshold to about 500 MeV 
(where the experimental data appear) there are no opened channels. 
Therefore, at the adequate representation of the amplitude, its 
continuation to the threshold is unique. 
%
%
\begin{table}[ht] 
\caption{Comparison of the $\pi\pi$ scattering length from various approaches.} 
\label{scatlength} 
\begin{center} 
\begin{ruledtabular} 
\begin{tabular}{lll} 
$a_1^1[10^{-3}m_{\pi^+}^{-3}]$ &{~~~~~~~~~References} & ~~~~~~~~~~Remarks \\ 
\hline 
$33.9\pm 2.02$ & This paper & Breit--Wigner analysis\\ 
$ 34 $ & \cite{NJL92} & Local NJL model \\ 
$ 37 $ & \cite{Osi06} & Non-local NJL model \\ 
$37.9\pm 0.5$ & \cite{Cap08} & Roy equations using ChPT \\ 
$39.6\pm 2.4$ & \cite{Kam03}& Roy equations \\ 
$38.4\pm 0.8$ & \cite{Pel05} & Forward dispersion relations 
\end{tabular} 
\end{ruledtabular} 
\end{center} 
\end{table} 
 
\section{Analysis of isoscalar-tensor sector } 
 
In analysis of the processes 
$\pi\pi\to\pi\pi,K\overline{K},\eta\eta$, we considered explicitly 
also the channel $(2\pi)(2\pi)$. Here it is impossible to use the 
uniformizing-variable method. Therefore, using the Le 
Couteur-Newton relations, we generate the resonance poles by some 
4-channel Breit-Wigner forms. The $d(k_1,k_2,k_3,k_4)$-function is 
taken as ~$d=d_B d_{res}$, where the resonance part is  
\begin{equation} 
d_{res}(s)=\prod_{r} 
\left[M_r^2-s-i\sum_{j=1}^4\rho_{rj}^5R_{rj}f_{rj}^2\right] 
\end{equation} 
with $\rho_{rj}=2k_j/\sqrt{M_r^2-4m_j^2}$ and $f_{rj}^2/M_r$ the 
partial width. The Blatt--Weisskopf barrier factor for a tensor 
particle is 
\begin{equation} 
R_{rj}=\frac{9+\frac{3}{4} 
(\sqrt{M_r^2-4m_j^2}~r_{rj})^2+\frac{1}{16}(\sqrt{M_r^2-4m_j^2}~r_{rj})^4} 
{9+\frac{3}{4}(\sqrt{s-4m_j^2}~r_{rj})^2+\frac{1}{16}(\sqrt{s-4m_j^2}~r_{rj})^4}, 
\end{equation} 
with radii of 0.943 fm for all resonances in all channels, except 
for $f_2(1270)$ and $f_2(1960)$ for which they are: for 
$f_2(1270)$,~ 1.498, 0.708, and 0.606 fm in the channels $\pi\pi$, 
$K\overline{K}$, and $\eta\eta$, respectively;  
for $f_2(1960)$, ~0.296 fm in the channel $K\overline{K}$. 
 
The background part has the form 
\begin{equation} 
d_B=\mbox{exp}\left[-i\sum_{n=1}^{3} 
\left(\frac{2k_n}{\sqrt{s}}\right)^5(a_n+ ib_n)\right]  
\end{equation} 
with 
\begin{eqnarray} 
a_1=\alpha_{11}+\frac{s-4m_K^2}{s}~\alpha_{12}~\theta(s-4m_K^2)+ 
\frac{s-s_v}{s}~\alpha_{10}~\theta(s-s_v)),\\ 
b_n=\beta_n+\frac{s-s_v}{s}~\gamma_n~\theta(s-s_v). 
\end{eqnarray} 
$s_v\approx2.274$ GeV$^2$ is a combined threshold of the channels 
$\eta\eta^{\prime}$, $\rho\rho$, and $\omega\omega$. 
 
The data for the $\pi\pi$ scattering are taken from an energy-independent 
analysis by Hyams {\it et al.}~\cite{Hya73}. The data for 
$\pi\pi\to K\overline{K},\eta\eta$ are taken from works \cite{Lin92}. 
 
We obtained a satisfactory description with ten resonances 
$f_2(1270)$, $f_2(1430)$, $f_2^{\prime}(1525)$, $f_2(1580)$, 
$f_2(1730)$, $f_2(1810)$, $f_2(1960)$, $f_2(2000)$, $f_2(2240)$,  
and $f_2(2410)$ (the total 
$\chi^2/\mbox{NDF}=161.147/(168-65)\approx 1.56$) and with eleven 
states adding one more resonance $f_2(2020)$ which is needed in 
the combined analysis of processes 
$p\overline{p}\to\pi\pi,\eta\eta,\eta\eta^\prime$ \cite{Ani05}. 
In our analysis, the description with eleven resonances is practically  
the same as that with ten resonances: the total 
$\chi^2/\mbox{NDF}=156.617/(168-69)\approx 1.58$. 
 
The obtained resonance parameters are shown in Table~\ref{tab:tensor_param} 
for the cases of ten and eleven states. 
%
%
\begin{table}[htb!] 
\caption{The resonance parameters in the tensor sector for ten and eleven  
states (in MeV).} 
\label{tab:tensor_param} 
\begin{center} 
\begin{ruledtabular} 
{\small 
\begin{tabular}{ccccccc} 
{State} & ~$M$~ & $f_{r1}$ & $f_{r2}$ & $f_{r3}$ & $f_{r4}$ & $\Gamma_{tot}$\\ 
\hline 
\multicolumn{7}{c}{ten states}\\ \hline 
{$f_2(1270)$} & 1275.3$\!\pm\!$1.8 & 470.8$\!\pm\!$5.4 & 201.5$\!\pm\!$11.4 
& 90.4$\!\pm\!$4.76 & 22.4$\!\pm\!$4.6 & $\approx$212\\ 
{$f_2(1430)$} & 1450.8$\!\pm\!$18.7 & 128.3$\!\pm\!$45.9 & 562.3$\!\pm\!$142 
& 32.7$\!\pm\!$18.4 & 8.2$\!\pm\!$65 & $>$230\\ 
{$f_2^{\prime}(1525)$} & 1535$\!\pm\!$8.6 & 28.6$\!\pm\!$8.3 & 253.8$\!\pm\!$78 
& 92.6$\!\pm\!$11.5 & 41.6$\!\pm\!$160 & $>$49\\ 
{$f_2(1565)$} & 1601.4$\!\pm\!$27.5 & 75.5$\!\pm\!$19.4 & 315$\!\pm\!$48.6 
& 388.9$\!\pm\!$27.7 & 127$\!\pm\!$199 & $>$170\\ 
{$f_2(1730)$} & 1723.4$\!\pm\!$5.7 & 78.8$\!\pm\!$43 & 289.5$\!\pm\!$62.4 
& 460.3$\!\pm\!$54.6 & 107.6$\!\pm\!$76.7 & $>$182\\ 
{$f_2(1810)$} & 1761.8$\!\pm\!$15.3 & 129.5$\!\pm\!$14.4 & 259$\!\pm\!$30.7 
& 469.7$\!\pm\!$22.5 & 90.3$\!\pm\!$90 & $>$177\\ 
{$f_2(1960)$} & 1962.8$\!\pm\!$29.3 & 132.6$\!\pm\!$22.4 & 333$\!\pm\!$61.3 
& 319$\!\pm\!$42.6 & 65.4$\!\pm\!$94 & $>$119\\ 
{$f_2(2000)$} & 2017$\!\pm\!$21.6 & 143.5$\!\pm\!$23.3 & 614$\!\pm\!$92.6 
& 58.8$\!\pm\!$24 & 450.4$\!\pm\!$221  & $>$299\\ 
{$f_2(2240)$} & 2207$\!\pm\!$44.8 & 136.4$\!\pm\!$32.2 & 551$\!\pm\!$149 
& 375$\!\pm\!$114 & 166.8$\!\pm\!$104 & $>$222\\ 
{$f_2(2410)$} & 2429$\!\pm\!$31.6 & 177$\!\pm\!$47.2 & 411$\!\pm\!$196.9 
& 4.5$\!\pm\!$70.8 &460.8$\!\pm\!$209 & $>$170\\ \hline 
\multicolumn{7}{c}{eleven states}\\ \hline 
{$f_2(1270)$} & 1276.3$\!\pm\!$1.8 & 468.9$\!\pm\!$5.5 & 201.6$\!\pm\!$11.6 
& 89.9$\!\pm\!$4.79 & 7.2$\!\pm\!$4.6 & $\approx$210.5\\ 
{$f_2(1430)$} & 1450.5$\!\pm\!$18.8 & 128.3$\!\pm\!$45.9 & 562.3$\!\pm\!$144 
& 32.7$\!\pm\!$18.6 & 8.2$\!\pm\!$63 & $>$230\\ 
{$f_2^{\prime}(1525)$} & 1534.7$\!\pm\!$8.6 & 28.5$\!\pm\!$8.5 & 253.9$\!\pm\!$79 
& 89.5$\!\pm\!$12.5 & 51.6$\!\pm\!$155 & $>$49.5\\ 
{$f_2(1565)$} & 1601.5$\!\pm\!$27.9 & 75.5$\!\pm\!$19.6 & 315$\!\pm\!$50.6 
& 388.9$\!\pm\!$28.6 & 127$\!\pm\!$190 & $>$170\\ 
{$f_2(1730)$} & 1719.8$\!\pm\!$6.2 & 78.8$\!\pm\!$43 & 289.5$\!\pm\!$62.6 
& 460.3$\!\pm\!$545. & 108.6$\!\pm\!$76. & $>$182.4\\ 
{$f_2(1810)$} & 1760$\!\pm\!$17.6 & 129.5$\!\pm\!$14.8 & 259$\!\pm\!$32. 
& 469.7$\!\pm\!$25.2 & 90.3$\!\pm\!$89.5 & $>$177.6\\ 
{$f_2(1960)$} & 1962.2$\!\pm\!$29.8 & 132.6$\!\pm\!$23.3 & 331$\!\pm\!$61.5 
& 319$\!\pm\!$42.8 & 62.4$\!\pm\!$91.3 & $>$118.6\\ 
{$f_2(2000)$} & 2006$\!\pm\!$22.7 & 155.7$\!\pm\!$24.4 & 169.5$\!\pm\!$95.3 
& 60.4$\!\pm\!$26.7 & 574.8$\!\pm\!$211  & $>$193\\ 
{$f_2(2020)$} & 2027$\!\pm\!$25.6 & 50.4$\!\pm\!$24.8 & 441$\!\pm\!$196.7 
& 58$\!\pm\!$50.8 & 128$\!\pm\!$190 & $>$107\\ 
{$f_2(2240)$} & 2202$\!\pm\!$45.4 & 133.4$\!\pm\!$32.6 & 545$\!\pm\!$150.4 
& 381$\!\pm\!$116 & 168.8$\!\pm\!$103 & $>$222\\ 
{$f_2(2410)$} & 2387$\!\pm\!$33.3 & 175$\!\pm\!$48.3 & 395$\!\pm\!$197.7 
& 24.5$\!\pm\!$68.5 & 462.8$\!\pm\!$211 & $>$168 
\end{tabular} 
} 
\end{ruledtabular} 
\end{center} 
\end{table} 
 
The background parameters for ten resonances are:  
$\alpha_{11}=-0.07805$, $\alpha_{12}=0.03445$, 
$\alpha_{10}=-0.2295$, $\beta_1=-0.0715$, $\gamma_1=-0.04165$, 
$\beta_2=-0.981$, $\gamma_2=0.736$, $\beta_3=-0.5309$, 
$\gamma_3=0.8223$; and for eleven resonances are:  
$\alpha_{11}=-0.0755$, $\alpha_{12}=0.0225$, 
$\alpha_{10}=-0.2344$, $\beta_1=-0.0782$, $\gamma_1=-0.05215$, 
$\beta_2=-0.985$, $\gamma_2=0.7494$, $\beta_3=-0.5162$, 
$\gamma_3=0.786$. 
 
\begin{figure}[htb] 
\includegraphics[width=0.84\textwidth]{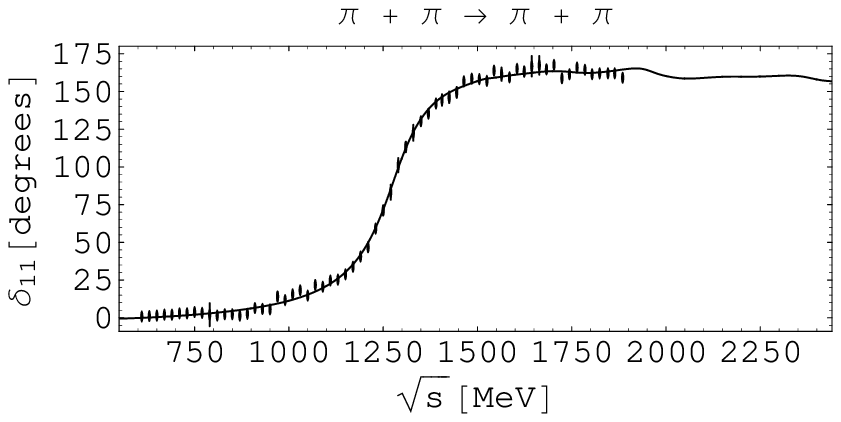} 
\includegraphics[width=0.84\textwidth]{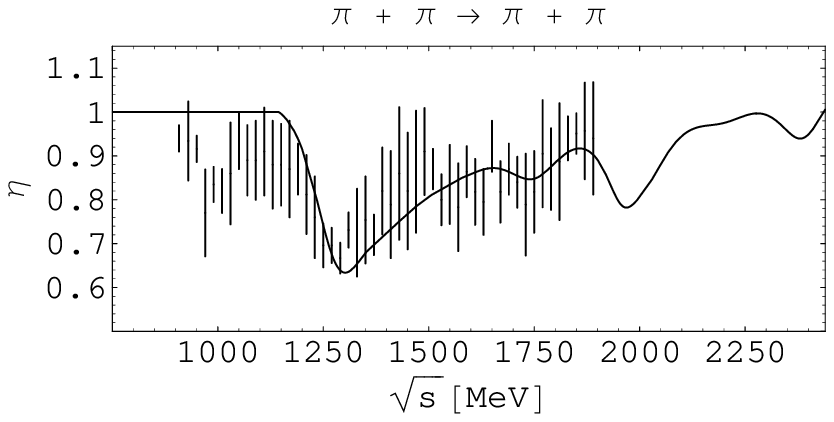} 
\caption{The phase shift and module of the $\pi\pi$-scattering 
$D$-wave $S$-matrix element.} \label{y12} 
\end{figure}

\begin{figure}[htb] 
\includegraphics[width=0.83\textwidth]{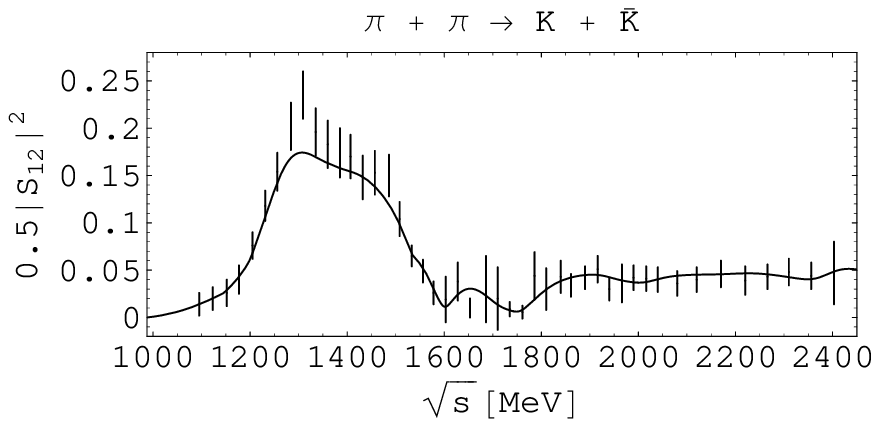} 
\includegraphics[width=0.83\textwidth]{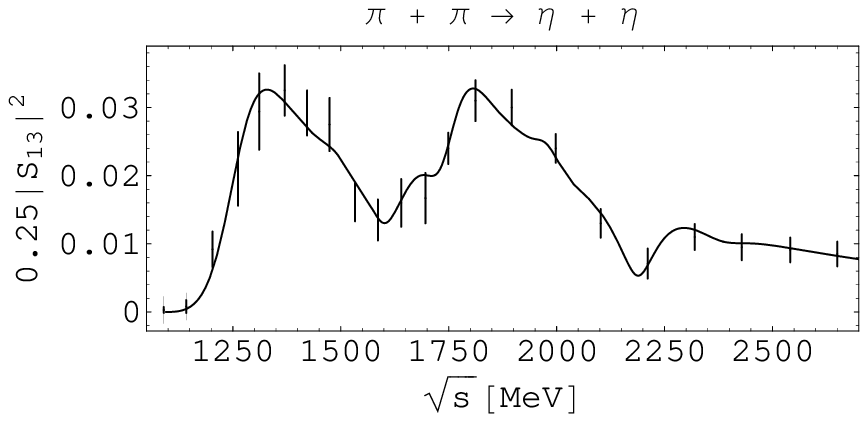} 
\caption{The squared modules of the $\pi\pi\to K\overline{K}$ 
(upper figure) and $\pi\pi\to\eta\eta$ (lower figure) $D$-wave 
$S$-matrix elements.} \label{y34} 
\end{figure}

In Figures \ref{y12} and \ref{y34} we show results of fitting to the data. 
 
\section{Spectroscopic implications from the analysis} 
 
In the combined model-independent analysis of data on the 
$\pi\pi\to\pi\pi,K\overline{K},\eta\eta,\eta\eta^\prime$ processes 
in the channel with $I^GJ^{PC}=0^+0^{++}$, an additional confirmation 
of the $\sigma$-meson with mass 835 MeV is obtained (the pole position 
on sheet II is $598-i583$ MeV). 
This value of mass corresponds most near to the one ($\sim 860$ MeV) of 
Ref. \cite{Tornqvist} and rather accords with prediction ($m_\sigma\approx 
m_\rho$) on the basis of mended symmetry by S.~Weinberg \cite{Wei90}. 
Note that our values of $E_r$ and $\Gamma_r$ for the $f_0(600)$-pole position 
are larger than those obtained in the dispersive analysis of data on 
only the $\pi\pi$ scattering, see Ref.~\cite{G-MKP08} and reference therein. 
 
Indication for $f_0(980)$ to be the $\eta\eta$ bound state is obtained. 
From the point of view of the quark structure, this is the 4-quark state. 
Maybe, this is consistent somehow with arguments in favour of the 4-quark 
nature of $f_0(980)$ \cite{20}. 
 
The ${f_0}(1370)$  and $f_0 (1710)$ have the dominant $s{\bar s}$ 
component. Conclusion about the ${f_0}(1370)$ agrees quite well 
with the one drawn by the Crystal Barrel Collaboration \cite{21} 
where the ${f_0}(1370)$ is identified 
as $\eta\eta$ resonance in the $\pi^0\eta\eta$ final state of the 
${\bar p}p$ annihilation at rest. Conclusion about the $f_0 
(1710)$ is quite consistent with the experimental facts that this 
state is observed in $\gamma\gamma\to K_S{\bar K}_S$ 
\cite{22} and not observed in $\gamma\gamma\to\pi^+\pi^-$ \cite{23}. 
 
As to the $f_0(1500)$, we suppose that it is practically the eighth  
component of octet mixed with the glueball being dominant in this state.  
Its biggest width among the enclosing states tells also in behalf of  
its glueball nature \cite{24}. 
 
We propose the following assignment of scalar mesons below 1.9 GeV 
to lower nonets, excluding the $f_0(980)$ as the $\eta\eta$ bound state.  
\underline{The lowest nonet}: the isovector $a_0(980)$, the isodoublet  
$K_0^*(900)$, and $f_0(600)$ and $f_0(1370)$ as mixtures of the eighth  
component of octet and the SU(3) singlet. Then the Gell-Mann--Okubo  
(GM-O) formula 
\begin{equation} 
 3m_{f_8}^2=4m_{K_0^*}^2-m_{a_0}^2\,, 
\end{equation} 
gives $m_{f_8}=872$ MeV ($m_\sigma=835\pm14$~MeV).  
In the relation for masses of nonet   
\begin{equation} 
m_\sigma+m_{f_0(1370)}=2m_{K_0^*}\,, 
\end{equation}  
the left-hand side is about 25 \% bigger than the right-hand one. 
 
\underline{The next nonet}: $a_0(1450)$, 
$K_0^*(1450)$, and $f_0(1500)$ and $f_0(1710)$. From 
the GM-O formula, we get $m_{f_8}\approx1450$ MeV. In the relation  
\begin{equation}  
m_{f_0(1500)}+m_{f_0(1710)}=2m_{K_0^*(1450)}\,, 
\end{equation} 
the left-hand side is about 12 \% bigger than the right-hand one. 
 
Now an adequate mixing scheme should be found. 
 
\underline{In the vector sector}, the obtained value of mass for  
the $\rho(770)$ is smaller in the model-independent approach, $769.3$ MeV,   
and a little bit bigger in the Breit--Wigner one, $777.69\pm0.32$ MeV,  
than the averaged value cited in the PDG tables \cite{PDG08},  
$775.49\pm0.34$ MeV.  
However, it also occurs in analysis of some reactions (see PDG tables).  
The obtained value of the total width in the first case ($146.6$ MeV)  
is in a good agreement  with the averaged PDG one ($149.4\pm1.0$ MeV)  
and it is a little bit bigger in the second case ($\approx154.3$ MeV)  
than the averaged PDG value, however, this is encountered also in  
other analyses (see PDG tables). 
Note that predicted widths of the $\rho(770)$ decays to the 
$4\pi$-modes are significantly larger than, {\it e.g.}, the ones 
evaluated in the chiral model of some mesons based on the hidden 
local symmetry added with the anomalous terms \cite{Ach05}. 
 
The first $\rho$-like meson has the mass 1257.8$\pm$11 MeV in the 
model-independent analysis and 1249.8$\pm$15.6 MeV in the 
Breit--Wigner one. These values differ significantly from the mass 
(1459$\pm$11~MeV) of the first $\rho$-like meson cited in the PDG 
tables. The $\rho(1250)$ was discussed actively some time 
ago \cite{25} and later the evidence for its existence was 
obtained in \cite{SB_NPA08,26}. 
 
If the $\rho(1250)$ is interpreted as the first radial excitation 
of the $1^+1^{--}$ $q{\bar q}$ state, then it lies down well on 
the corresponding linear trajectory with an universal slope on the 
$(n,M^2)$ plane (n is the radial quantum number of the $q{\bar q}$ 
state)\cite{27}, whereas the $\rho(1450)$ turns out to 
be considerably higher than this trajectory. The $\rho(1250)$ and 
the isodoublet $K^*(1410)$ are well located to the octet of the first 
radial excitations. The mass of the latter should be by about 150 
MeV larger than the mass of the former. Then the GM-O formula 
\begin{equation} 
3m_{\omega_8^\prime}^2=4m_{{K^*}^\prime}^2-m_{\rho^\prime}^2 
\end{equation} 
gives $m_{\omega_8^\prime}=1460$ MeV, that is fairly 
good compatible with the mass of the first $\omega$-like meson 
$\omega(1420)$, for which one obtains the values in range 
1350-1460 MeV (see PDG tables). 
 
Existence of the $\rho(1450)$ (along with $\rho(1250)$) does not 
contradict to the data. In the $q{\bar q}$ picture, it might be 
the first $^3D_1$ state with, possibly, the isodoublet $K^*(1680)$ 
in the corresponding octet. From the GM-O formula, we should 
obtain the value 1750 MeV for the mass of the eighth component of 
this octet. This corresponds to one of the observations of the 
second $\omega$-like meson with masses from 1606 to 1840 MeV that 
is cited in the PDG tables under the $\omega(1650)$. 
 
The third $\rho$-like meson has the mass about 1600 MeV rather 
than 1720 MeV cited in the PDG tables \cite{PDG08}. 
 
As to the $\rho(1900)$, in this energy region there are 
practically no data on the $P$-wave of $\pi\pi$ scattering. The 
model-independent analysis testifies in favour of existence of 
this state, whereas the Breit--Wigner analysis gives the same 
description with and without the $\rho(1900)$. 
 
The suggested picture for the first two $\rho$-like mesons is 
consistent with predictions of the quark model \cite{28}. 
In Ref.~\cite{29} the discussed mass spectrum for 
radially excited $\rho$ and $K^*$ mesons was obtained using 
rather simple mass operator. 
If the existence of the $\rho(1250)$ is confirmed, some 
quark potential models, {\it e.g.}, in Ref.~\cite{30}, will require 
substantial revisions, 
because the first $\rho$-like meson is usually predicted about 200 
MeV higher than this state. To the point, the first $K^*$-like 
meson is obtained in the indicated quark model at 1580 MeV, 
whereas the corresponding very well established resonance has the 
mass of only 1410 MeV. 
 
\underline{In the tensor sector}, we carried out two analysis -- without  
and with the $f_2(2020)$. 
We do not obtain $f_2(1640)$, $f_2(1910)$ and $f_2(2150)$, 
however, we see $f_2(1450)$ and $f_2(1730)$ which are related to the 
statistically-valued experimental points. 
 
Usually one assigns the states $f_2(1270)$ and $f_2^{\prime}(1525)$  
to the ground tensor nonet. To the second nonet, one could assign 
$f_2(1600)$ and $f_2(1760)$ though for now the isodoublet member is 
not discovered. If $a_2(1730)$ is the isovector of this octet and if 
$f_2(1600)$ is almost its eighth component, then, from the GM-O 
formula, we expect this isodoublet mass at about 1633 MeV. 
Then the relation for masses of nonet would be fulfilled with  
a 3\% accuracy. Karnaukhov {\it et al.} \cite{31} observed the strange  
isodoublet with yet indefinite remaining quantum numbers and with mass 
$1629\pm7$ MeV in the mode $K_s^0\pi^+\pi^-$. This state might be 
the tensor isodoublet of the second nonet. 
 
The states $f_2(1963)$ and $f_2(2207)$ together with the 
isodoublet $K_2^*(1980)$ could be put into the third nonet. Then 
in the relation for masses of nonet 
\begin{equation} 
M_{f_2(1963)}+M_{f_2(2207)}=2M_{K_2^*(1980)}, 
\end{equation} 
the left-hand side is only 5.3 \% bigger than the right-hand one.  
If one consider $f_2(1963)$ as the eighth component of octet, the 
GM-O formula 
\begin{equation} 
M_{a_2}^2=4M_{K_2^*(1980)}^2-3M_{f_2(1963)}^2 
\end{equation} 
gives $M_{a_2}=2030$ MeV. This value coincides with the one for 
$a_2$-meson obtained in works \cite{32}.  
This state is interpreted as a second radial excitation of the 
$1^-2^{++}$-state on the basis of consideration of the $a_2$ 
trajectory on the $(n,M^2)$ plane \cite{Ani05}. 
 
As to $f_2(2000)$, the presence of the $f_2(2020)$ in the analysis 
with eleven resonances helps to interpret $f_2(2000)$ as the 
glueball. In the case of ten resonances, the ratio of the $\pi\pi$ 
and $\eta\eta$ widths is in the limits obtained in 
Ref.~\cite{Ani05}  for the tensor glueball on the basis of 
the 1/N-expansion rules. However, the $K\overline{K}$ width is too 
large for the glueball. At practically the same description of 
processes with the consideration of eleven resonances as in the 
case of ten, their parameters have varied a little, except for the 
ones for $f_2(2000)$ and $f_2(2410)$. Mass of the latter has 
decreased by about 40 MeV. As to $f_2(2000)$, its $K\overline{K}$ 
width has changed significantly. Now all the obtained ratios of 
the partial widths are in the limits corresponding to the 
glueball. 
 
The question of interpretation of the $f_2(1450)$, $f_2(1730)$, 
$f_2(2020)$ and $f_2(2410)$ is open.

\end{document}